\documentclass[5p]{elsarticle}
\usepackage{tensor}
\usepackage{mathrsfs}
\usepackage{amsmath}
\usepackage{aas}
\usepackage{amssymb}
\usepackage{standalone}
\usepackage{wasysym}

\newcommand{\mytitle}[0]{ngravs: Distinct gravitational interactions in \textsc{gadget}-2}
\newcommand{\myauthor}[0]{K.~A.~S.~Croker}
\newcommand{\dd}[0]{\mathrm{d} }
\usepackage[pdfauthor={\myauthor},pdftitle={\mytitle}]{hyperref}

\journal{Computer Physics Communications}
\biboptions{sort&compress}
\begin{document}

\begin{frontmatter}

\title{\mytitle}
\author[kasc]{K. A. S. Croker}
\address[kasc]{Department of Physics and Astronomy, University of Hawai`i at M\=anoa, 2505 Correa Road, Honolulu, HI., USA}
\ead{kcroker@phys.hawaii.edu}

\begin{abstract}
We discuss an extension of the massively parallel cosmological
simulation code \textsc{gadget}-2, which now enables investigation of
multiple and distinct gravitational force laws, provided they are
dominated by a constant scaling of the Newtonian force.  In addition
to simplifying investigations of a universally modified force law, the
\emph{ngravs} extension provides a foundation for state-of-the-art
collisionless cosmological simulations of exotic gravitational
scenarios.  We briefly review the algorithms used by
\textsc{gadget}-2, and present our extension to multiple gravities,
highlighting additional features that facilitate consideration of
exotic force laws.  We discuss the accuracy and performance of the
\emph{ngravs} extension, both internally and with an unaltered
\textsc{gadget}-2, in the relevant operational modes.  The
\emph{ngravs} extension is publicly released to the research
community.
\end{abstract}

\begin{keyword}
numerical methods, gravitation, cosmology, n-body
\end{keyword}

\end{frontmatter}

%
\section*{PROGRAM SUMMARY}

\begin{small}
\noindent
{\em Manuscript Title:} ngravs: Distinct gravitational interactions in \textsc{gadget}-2                                       \\
{\em Authors:} K. A. S. Croker                                               \\
{\em Program Title:} Gadget-2.0.7-ngravs                                         \\
{\em Journal Reference:}                                      \\
{\em Catalogue identifier:}                                   \\
{\em Licensing provisions:} GNU General Public License v2                                   \\
{\em Programming language:}  C                                 \\
{\em Computer:} Commodity                                              \\
{\em Operating system:} Unix                                      \\
{\em RAM:} 256MB+                                              \\
{\em Number of processors used:} MPI                              \\
{\em Keywords:} numerical methods, gravitation, cosmology, n-body  \\
{\em Classification:} 1.9 Cosmology, 4.12 Other Numerical Methods                                         \\
{\em External routines/libraries:} Same requirements as \textsc{gadget}-2                           \\
{\em Nature of problem:}\\
$N$-body cosmological codes are traditionally designed to investigate
a single gravitating species interacting via the Newtonian force law.
There exist viable extensions to General Relativity[1], however, which
predict weak-field, slow-motion limits featuring distinct
gravitational force laws between distinct particle species.  To enable
investigation and constraint of these theories with available
astrophysical data, a necessary first step is to extend an $N$-body
simulator to handle distinct gravitating species.  \\ {\em Solution
  method:}\\
The massively parallel Barnes-Hut tree, Fast Fourier Transform, and
sorting routines of the versatile and well vetted $N$-body
simulator[2] \textsc{gadget}-2 were extended to support $D$ distinct
gravitationally interacting species.  The tree implementation now
vectorizes over each species' monopole masses and positions, the
Fourier routines now handle active and passive gravitational masses
separately, and the sorting routines now group all particle data by
type.  The appropriate TreePM adjusted forces are computed via FFT and
tabulated before runtime.  An additional file was introduced allowing
the user to specify all $D^2$ gravitational interactions: real space,
Fourier space, and lattice summation corrections.  To improve monopole
approximations in scenarios where the scale of the gravitational
interaction depends on the mass itself, an optional tracking of the
number of bodies contributing to any particular monopole approximation
has been written.  \\
{\em Restrictions:}\\
Mesh methods with non-periodic boundary conditions have been disabled.
Force laws with mass dependent scale lengths are not amenable to the
implemented Fourier methods (or even the traditional[3] Fourier
approach).  Nodes containing highly heterogeneous collections of
particles with different mass dependent scale lengths may not be
well-approximated, even with the additional tracking introduced.  The
collisional ``gas'' species can only interact via a single
gravitational force law.  \\ {\em Unusual features:}\\
The extension allows simultaneous consideration of at most six distinct
central forces, where each is a sum of bounded, monotonic, gradients
of radially modulated Newtonian potentials.  This will serve as a
common platform for model-dependent adjustments to the cosmological
background evolution. \\ {\em Additional comments:}\\
Data file format is identical to that of \textsc{gadget}-2.
Configuration file format is unchanged, save for the addition of
required bindings between particle species and gravitational type.  To
install ngravs, first install \textsc{gadget}-2 (available at
http://www.mpa-garching.mpg.de/gadget/ ), then replace the contents of
the Gadget2 subdirectory with the files included in the provided
tarball.  Alternatively, one can clone the github repository
https://github.com/kcroker/Gadget-2.0.7-ngravs and provide one's own
configuration and initial data files.   \\ {\em Running time:}\\
Typical
running times are $\lesssim 2D\times$ those of
\textsc{gadget}-2, where $D$ is an integer between 1 and 6.  \\

\end{small}

%

\section{Introduction}
At present, there is substantial evidence that most of the mass and
energy within our universe is non-luminous.  Big bang nucleosynthesis
and baryon acoustic oscillations strongly
constrain\cite{DodelsonCosmology} the fraction of non-luminous matter
with respect to the luminous component from the radiation dominated
era onward.  Yet, while they and other evidence, such as the Bullet
Cluster\cite{Clowe2006}, strongly suggest that the dark fraction can
be well-approximated by pressureless ideal fluid, the composition and
precise distribution of this dark matter is far from clear.  While
numerous particle dark matter candidates\cite{HooperReview} are
theoretically popular at present, conflicting exclusions from possible
detections and increasingly stringent constraints from lack of direct
detection\cite{LUX15} at cosmologically desirable mass scales continue
to motivate investigation in novel directions.

With the growing wealth of high-precision astrophysical
data\cite{SDSS2012, WMAP9, XVIPlanck2013} and the absence of a
``bottom-up'' understanding of dark matter, $N$-body
methods\cite{Hockney81} have become essential for comparison against
highly non-linear theoretical predictions in structure formation over
many decades of spatio-temporal scale.  In the past twenty years, the
sophistication of such cosmological simulations both in physical scope
and technical implementation has undergone unprecedented
growth\cite{BaglaNbody, Millenium2005Natur, Illustris}.  While
significant literature exists on the use of $N$-body methods to
explore standard structure formation scenarios, there exist proposed
exotic scenarios that are also amenable to $N$-body methods.  In a
particularly intriguing case, Hohmann \& Wohlfarth\cite{Hohmann2010vt}
proposed to model Dark Energy as a \emph{repulsive} gravitational
interaction between matter species belonging to distinct copies of the
Standard Model.  This model should make very strong predictions for
weak-lensing observations and could be readily adapted to the dark
matter problem.  This motivates the extension of existing numerical
simulation tools to investigate and constrain more general
cosmological models.

One well-established and versatile simulation tool is
\textsc{gadget}-2: a massively parallel code with extensive memory and
speed optimization employed both algorithmically and
architecturally\cite{Springel2005}.  \textsc{gadget}-2 has been
extended to consider three distinct gravitationally interacting
species by Baldi~et.~al.~\cite{Baldi2010} to enable investigation of
coupled dark-energy cosmologies.  Unfortunately, their extension is
focused on particular cosmological models and not publicly available.

In the following, we discuss our independent and augmented
implementation of $D$ distinct gravitationally interacting species in
\textsc{gadget}-2.  To facilitate investigation of models amenable to
multi-species treatments by individual researchers, our publicly
released \emph{ngravs} extension permits convenient definition of
$D^2$ distinct gravitational force laws and optionally provides
additional data which can be used to improve force accuracy under
certain scenarios.  Our primary aim is to facilitate large scale
structure investigations of multi-species models, with considerable
freedom in the precise form of the gravitational force laws.  Our
initial use of \emph{ngravs} will be the constraint of such models
through predicted galaxy power spectra.

We assume the reader is familiar with the goals, construction, and
operation of modern $N$-body codes.  Throughout this paper, $D$ will
always refer to the number of distinct gravitationally interacting
species, while $N$ will refer to the number of bodies considered in
any particular simulation.  Units will be such that $G\equiv c\equiv
1$.  Stock will refer to the unaltered \textsc{gadget}-2.0.7 code,
while \emph{ngravs} will refer to our augmented version of this same
code.

\section{Implementation}
The algorithms employed by \textsc{gadget}-2 to compute collisionless
forces are a Barnes-Hut tree walk and, optionally, a particle mesh
(PM) computation.  To construct the tree, the simulation volume is
recursively halved until each particle resides within its own leaf.
Interparticle forces are then computed for any specific particle by
recursive traversal of the tree, halted when a monopole approximation
of the force from all deeper branches satisfies a user-specified force
accuracy ``opening criterion.''  PM forces are computed by
interpolating particle positions to a mass density defined on a
regular grid, performing a Fourier transformation, convolving with the
$k$-space Greens' function, and inverting the transform.  The
resulting potential is then interpolated back to forces at all
particle positions.  The tree algorithm may be used exclusively, or
can be combined with the PM algorithm in a hybrid arrangement (TreePM)
where the tree is used to compute short range forces, and the PM
algorithm used to rapidly compute distant contributions.  In TreePM
mode, the $k$-space Greens' function is Gaussian filtered, and the
corresponding truncated short range force is summed with a spatially
restricted Tree walk.  The \textsc{gadget}-2 code is engineered so
that tree and PM computations are nearly decoupled from the more
intricate collisional force computations, time integration,
parallelization decomposition, and IO routines.  This permits a
targeted and straight-forward extension to multiple gravitational
interactions.
\begin{table}
\caption{\label{tbl:runtimes}Time averaged $D=1$ and $D=3$
  \emph{ngravs} tree computation performance with Newtonian
  interactions, normalized to stock runtimes.  $\dot{N}$ represents
  particles processed per second, adjusted for interprocess
  communication delays.  Subscript labels indicate stock (stc) or
  \emph{ngravs} ($D=1,3$).  Note that the purely collisional gas
  sphere cannot be tested for $D > 1$ (see \S~\ref{sec:limitations}).}
\begin{center}
{\renewcommand{\arraystretch}{1.1}
\begin{tabular}{r|cc}\hline
Test case & $\left<\dot{N}_{D=1}\right> / \left<\dot{N}_\mathrm{stc}\right>$ & $\left<\dot{N}_{D=3}\right> / \left<\dot{N}_\mathrm{stc}\right>$\\
\hline
Galaxy collision & $0.706$ & $0.334$ \\
$\Lambda$CDM gas & $0.779$ & $0.405$ \\
Gas sphere & $0.759$ & -\\
\hline
\end{tabular}}
\end{center}
\end{table}%

\textsc{gadget}-2 implements 6 distinct particle types, including a
``gas'' type which features additional force computations from
collisional dynamics.  Let $1 \leq D \leq 6$ be the number of distinct
gravitational species.  For particle types $i, j$, the following map
is established between $0 \leq i, j \leq K \equiv D-1$ generalizing
Newton's law of gravitation
\begin{align}
\label{eqn:map}
\vec{F}(m_i, M_j, r, N_\perp) = -\frac{m_i M_j}{r^2}\hat{r} \longrightarrow f_{ij}(r, N_\perp) \hat{r} 
\end{align}%
Here each $f_{ij}$ is dominated by a constant scaling of the Newtonian
force, with specific forms detailed in \S\ref{sec:specific_forms}.
They depend on the separation of masses $r$, the active mass $M_j$,
the passive mass $m_i$, and the number of source particles $N_\perp$
contributing to the monopole approximation employed by the Barnes-Hut
tree algorithm.  In the direct force case, $N_\perp \equiv 1$.

As the standard Newtonian force diverges as $r\to 0$, in order to
maintain numerical accuracy under reasonable timesteps,
\textsc{gadget}-2 artificially smooths to zero the gravitational
interaction below some (type dependent) length scale.  In general,
$D^2$ distinct ``softenings'' analogous to (\ref{eqn:map}) must also be
specified.  We have implemented the above mappings through function
pointers, enabling much model-dependent code to be conveniently
populated within a single location.  Thus, at the user's option,
tables for the gravitational potential, $\vec{k}$-space Greens'
functions, and lattice sum corrections may also be specified as
desired.

Pure tree computations have been extended to both periodic and
non-periodic modes, while TreePM computations have been extended only
to periodic mode at present, removing the possibility of secondary PM
``zoom'' simulations from the \emph{ngravs} extension.  For
justification of this design choice, we direct the reader to
\S~\ref{sec:limitations}.

Performance comparisons with respect to stock can be found in
Table~\ref{tbl:runtimes}, where we find an $\sim$40\% increased
runtime due to additional overhead within the tree algorithm.  It
should be noted that the simulation specific performance of
\emph{ngravs}, apart from this constant scaling, is unchanged from
that of stock.  Thus, in $D=1$ cosmological scenarios, the
\emph{ngravs} extension enables convenient investigation of a single,
globally modified force law; if one can afford modestly longer
runtimes.  More importantly, the structures containing data on all $N$
simulation particles are unchanged and so the favorable memory storage
requirements of \textsc{gadget}-2 are maintained.

\subsection{Specific forms of $f_{ij}$} 
\label{sec:specific_forms}
The Barnes-Hut algorithm makes assumptions about the nature of the
force law which must be honored to maintain accuracy of the
algorithm.  This is not a serious impediment, as arbitrary forces are
not relevant for cosmological investigations.  We now develop
properties of the $f_{ij}$ which will permit investigation of a very
wide range of possible scenarios, while simultaneously maintaining the
established force accuracies of \textsc{gadget}-2.

Investigations of the gravitational force between baryonic matter
strongly constrain this interaction to an inverse-square law
(ISL)\cite{Adelberger03}.  Analogous constraint for dark matter,
however, must presently come from large scale astrophysical data.
This leaves margin for speculation and motivates modification to the
gravitational interaction of the dark component.  Two popular
deviations from the \emph{baryonic} ISL, in the notation of
\cite{Adelberger03}, are the Yukawa-like
\begin{equation}
  \vec{F}(r) = GMm\frac{\dd}{\dd r}\frac{1}{r}\left[C + \alpha \exp\left(-r\over\lambda\right)\right]\hat{r}
\label{eqn:yukawa}
\end{equation}
and power-law
\begin{equation}
  \vec{F}(r) = GMm\frac{\dd}{\dd r}\frac{1}{r}\left[C + \alpha\left(\frac{r_0}{r}\right)^{M-1}\right]\hat{r}.
\label{eqn:powerlaw}
\end{equation}
We introduce the parameter $C\in\{0,1\}$ to permit consideration of a
``pure Yukawa'' law as could present in massive gravity
theories\cite{Hinterbichler12} or a harder power law, which would
follow from dimensional considerations if point-like dark matter
interactions were well-approximated by a Poisson law in greater than
$3$ non-compact spatial dimensions.  Note that
Equations~(\ref{eqn:yukawa}) with $C\equiv 1$ and (\ref{eqn:powerlaw})
strengthen the force law at small scales.

One may also consider the usual Newtonian force, but construct
``effective'' point-like force laws for extended mass distributions
which remain rigid on dynamical timescales.  An example could be a
dark matter ``particle'' sourced by a non-pointlike mass density
$\rho(r)$, stable on timescales relevant to simulation length.  Such
objects could be used to reduce particle count in a simulation, or to
investigate novel approaches to the missing satellites\cite{Klypin99}
and ``core cusp''\cite{de2009core} problems.  These objects would
interact with the usual baryonic matter via the following force
\begin{equation}
\vec{F}(r) = GMm\frac{\dd}{\dd r}\left[\frac{1}{r}\int_0^{r} \rho(r')
  r'^2\dd r'\right]\hat{r}
\label{eqn:rigid}
\end{equation} 
where the integral must remain finite as $r\to\infty$.  Note that, in
contrast to the above force laws with $C=1$,
Equation~(\ref{eqn:rigid}) diminishes the force law at all scales.

Equations (\ref{eqn:yukawa}), (\ref{eqn:powerlaw}), and
(\ref{eqn:rigid}) all take the form of the gradient of a modulated
Newtonian potential
\begin{equation}
\vec{F}_{ij}(r) = m\frac{\dd}{\dd r} {M s_{ij}(r)\over r}\hat{r} \equiv
-m\frac{\dd}{\dd r}V_{ij}(r)\hat{r}
\end{equation}
where $M$ is the active gravitational mass, $s_{ij}(r)$ is bounded,
positive, and $\lim_{r\to\infty} s_{ij}(r)$ monotonically approaches a
constant value.  In the following discussion, we restrict our
consideration to this class of force laws.  We emphasize that $M$ and
must appear as a multiplier in order to maintain the superposition
required by both the Tree and PM methods.  It is assumed that the
Equivalence Principle holds\cite{WillConfrontation}, and so $m$ must
also appear as a multiplier.  We will now drop this passive/inertial
mass and consider the accelerations $a_{ij}$ when convenient.

For cosmological simulations of metric theories of gravity, the
evolution of the scale factor is determined by the full field
equations.  Contributions to the weak field equations atop this
Robertson-Walker background are then determined by first order
perturbation theory.  While one can now investigate significantly more
general force laws below the horizon scale, if the background
expansion is significantly altered from the Friedmann equations,
additional adjustment to the timestep routines may be required to
obtain meaningful numerical results on cosmological scales.  Such
modifications are beyond the scope of the present work.

\subsubsection{Timesteps}
The determination of a suitable timestep for the numerical integration
is a subtle problem\cite{Gadget1}.  On the one hand, it is desirable
to use the largest possible timestep to speed the simulation, but one
must do so while maintaining force accuracy.  In general, the larger
the force, the smaller the required timestep.  Though there are many
approaches to determining a timestep\cite{dehnen2011n}, most require
knowledge of higher derivatives and would require significant
departure from the existing \textsc{gadget}-2 code.  We thus implement
a minimal departure from the algorithm of \textsc{gadget}-2.  This
guarantees the established force accuracy of \textsc{gadget}-2 and
facilitates comparision with studies performed with stock.

In the simplest scenario, \textsc{gadget}-2 maintains sympletic time
evolution by synchronously stepping each particle by
\begin{equation}
\Delta t_\mathrm{grav} = \min\left[ \Delta t_\mathrm{max}, \left(
  \frac{2 \eta \epsilon}{\left|\vec{a}\right|}\right)^{1/2}\right]
\label{eqn:timestep}
\end{equation}
with $\eta$ a user-specified dimensionless accuracy, $\vec{a}$ the
particle's acceleration, and $\epsilon$ a user-specified gravitational
softening length.  The maximum timestep $\Delta t_\mathrm{max}$ is
either specified by the user, or determined in cosmological
simulations by enforcing that rms displacement should be well below
mean particle separation.  Though the nonphysical $\epsilon$ enters
the timestep computation, this timestep has been shown to be robust
for Newtonian interactions through numerical
investigations\cite{Power2003}.
 
To \emph{guarantee} the already established force accuracy of
\textsc{gadget}-2 without alteration of the time integration
algorithms, we note that the permissible $f_{ij}$ exhibit an
$A_\mathrm{max}$ such that the accelerations $a_{ij}$ are bounded by
$A_\mathrm{max}/r^2$, provided one remains outside of an
appropriately defined softening scale.  Thus, we may modify
Equation~\ref{eqn:timestep} by a constant scaling
\begin{equation}
\Delta t_\mathrm{grav} = \min\left[ \Delta t_\mathrm{max}, \left(
  \frac{2 \eta \epsilon}{A_\mathrm{max}\left|\vec{a}\right|}\right)^{1/2}\right].
\label{eqn:our_timestep}
\end{equation}
This works because any acceleration smaller than that due
to this scaled Newtonian force will produce smaller
adjustments to any trajectory per unit time, and thus will be tracked
to greater precision by the existing timestepping algorithms.

\begin{figure}
\centering
\caption{\label{fig:Dvar_runtimes}Average particles processed per
  second $\dot{N}$ given $D$ species, normalized to $D = 1$.  Curves
  are separate one-parameter fits $cx^{-1}$ to the data.  Note the
  decrease proportional to $D^{-1}$ as expected in both pure tree and
  TreePM modes.  Rather large error bars for the TreePM scenario have
  been omitted for clarity as the relevant quantity is average
  performance.} 
\includegraphics{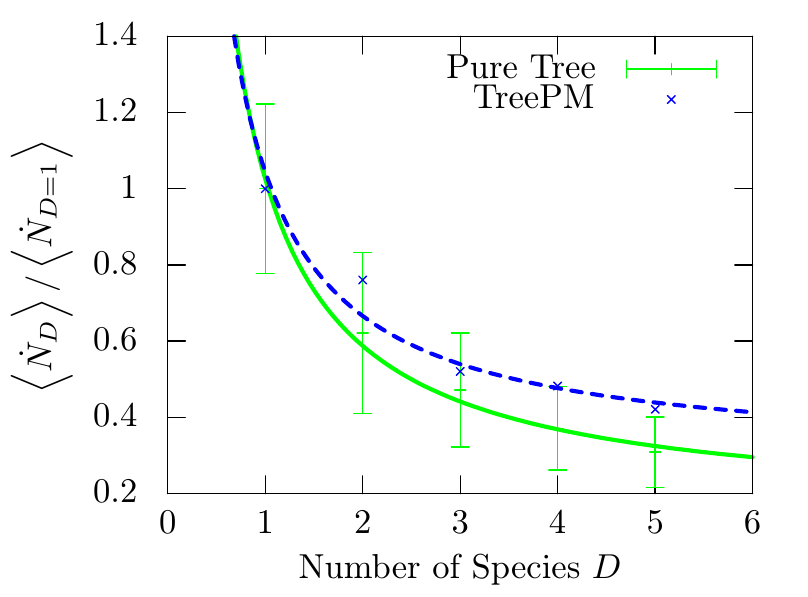}
\end{figure}%

\subsection{Tree forces}
In keeping with our approach, the existing tree routines were extended
simply by vectorizing over the monopole mass centers and velocities.
This additional data increases tree memory consumption by $\sim
0.3(D-1)$ from stock.  In order to accommodate more exotic force laws
for which the interaction scale is related to the active gravitational
mass in more complicated ways, the tree structure and construction
were augmented to optionally track $N_\perp$ for all contributing
types.  This quantity can then be used to suitably correct computation
of the monopole moment.

During force computation, tree walks may be significantly optimized by
suitable choice of opening criteria.  For collisionless force
computations in stock, the tree is traversed at most twice, once for
the usual particle-particle interaction, and once again for any
periodic correction.  The latter walk can proceed more efficiently, as
the opening criteria is significantly different: corrections to near
particles are very small.  We considered performing $D$ separate tree
walks, but stock employs a very effective relative opening criteria
during the tree walk, where the acceleration previously computed is
compared to a Newtonian estimate of the new acceleration to determine
whether to traverse the branch.  These previous accelerations,
however, are stored on a per-particle basis and the amount of memory
required to store $D-1$ additional accelerations produced an
unacceptable increase in memory consumption.  It was decided that,
since many alternative force-laws cannot deviate too strongly from the
Newtonian force, any gains in speed due to decreased depth within the
tree would not be offset by the additional memory requirement.
Instead, we vectorized within the single tree walk that
\textsc{gadget}-2 already performs, and continue to employ the
Newtonian relative opening criteria.  This opening criteria is
conservative, provided that the alternative force-laws are dominated
by the usual Newtonian interaction.  This results in slightly improved
force errors that can be easily understood: in a limit where the mass
distribution is characterized by $N$ distinct monopoles, one simply
regenerates the exact force. Overall, the tree walk runtime increases
by a factor of $D$.  Indeed such behavior is found in
Figure~\ref{fig:Dvar_runtimes}, where the average number of particles
processed per second decreases as $1/D$.

\subsection{Lattice corrections}
Stock periodic computations may optionally proceed in pure tree mode
with the method of Ewald summation\cite{Hernquist1991}.  The Ewald
technique is a specific application, to the ISL, of methods designed
to transform poorly convergent sums of periodic images in $r$-space
into rapidly convergent sums in $k$-space.  These methods belong to
the broader topic of lattice sums\cite{GlasserSums} which are, in
general, challenging to compute.  In terms of error functions,
computation can be reduced to quadrature\cite{Johnson07}, with
the usual caveats that apply to numerical integration.  \emph{ngravs}
can tabulate given corrections from an infinite image lattice for each
of $D^2$ direct forces, and interpolate to actual particle positions
in the same manner as stock.  In practice, the lattice correction is
usually specified as a consistency check against the Fourier
computations, and not used in actual simulations.

\subsection{PM forces}
In order to minimize surface to volume ratio, stock does not attempt
to overlap local PM computation with local particle distribution.
Instead, density data is exchanged between all parallel processes
according to an optimal slab decomposition determined by the Fast
Fourier Transform (FFT) routines\cite{FFTW05}, and the resulting
potential is exchanged back.  Since the identities of all the
particles which contribute to any given slab are unknown, the PM
routine must iterate $D^2$ times (instead of $D(D + 1)/2$ and
exploiting symmetry), so that each gravitational type can be both the
passive and active gravitational mass.  Though more sparse, the
exchanged data is of the same dimension and the runtime increases by a
factor of $D^2$.  This performance degradation is of little concern,
however, as the FFT runtime continues to be heavily
subdominant to that of the Tree algorithm, as is clear in
Figure~\ref{fig:Dvar_runtimes}.

In \textsc{gadget}-2, one may optionally 
enable Peano-Hilbert sorting of particle data on each local processor.
This was found by Springel\cite{Springel2005} to often give
substantial (but architecture dependent) improvements in runtime as
spatial proximity translates to memory proximity.  To enable
processing of the entire local particle content with only a single
traversal of the data, if $D>1$ and TreePM mode is enabled, we have
implemented an additional sort by gravitational type before the
Peano-Hilbert sort.  Subsequently, each gravitational type is then
Peano-Hilbert sub-sorted.  As is done in stock, both sorts proceed so
that only one reordering of the particle data memory is required.  For
compatibility with the collisional code of \textsc{gadget}-2,
collisional particles must be mapped to gravitational type zero.

\subsubsection{TreePM short-range forces}
In the hybrid TreePM mode, stock applies a Gaussian low-pass filter to
the $k$-space Newtonian potential.  This permits highly accurate and
rapid computation of a long-range Coloumb force with the PM algorithm.
On spatial scales near a user-specified $a_{smth}$ number of mesh
cells, the short-range force is calculated by smoothly transitioning
to a partial tree walk using a suitably adjusted potential
\begin{equation}
\phi^{\mathrm{short}}(r) \equiv \mathscr{F}^{-3}\left\{\phi_k\left[1 - \exp\left(-k^2r_s^2\right)\right] \right\}
\end{equation}
and taking a radial derivative, where $\mathscr{F}^{-3}$ denotes the
3D inverse Fourier transform, and $r_s \equiv 2\pi a_{smth}/L$.  It is
a convenient coincidence that, in the Newtonian case, this Fourier
transform yields $\phi^{\mathrm{short}}(r) =
\phi(r)\mathrm{erfc}(r/2r_s)$.  Stock samples a user-specified number
$N_\mathrm{TAB}$ of this factor and its derivative, and then
multiplies the computed tree potential and force by these respective
modulations.  For the case of a general force-law, we consider instead
\begin{equation}
\phi_{ij}^{\mathrm{short}}(r) = \phi_{ij}(r) - \frac{2\pi}{r}\int_0^r\mathscr{F}_{r'}^{-1}\left\{\bar{\phi}_{ij}(k)\exp(-k^2r_s^2)\right\}~\dd r'
\label{eqn:short_pot}
\end{equation}
where we have performed the angular integrations of the transform.  We
have also well-conditioned the computation by taking a derivative with
respect to $r$, and writing
\begin{align}
\bar{\phi}_{ij}(k) \equiv k^2\phi_{ij}(k)
\end{align}
which is just $\phi_{ij}(k)$ normalized by the Newtonian Greens'
function.  This lessens the severity of singularities in
$\phi_{ij}(k)$ before computation due to our prior constraint of the
$f_{ij}$.

Subtractions, such as in Equation~\ref{eqn:short_pot}, which involve
two values very near unity can suffer from loss of precision.  We
note, however, that the Fourier integrand of
Equation~\ref{eqn:short_pot} is effectively band-limited, so its
values may be rapidly and precisely (to near machine precision)
computed during initialization of \emph{ngravs} by an inverse FFT.
The acceleration is found from differentiation of
Equation~\ref{eqn:short_pot} with respect to $r$
\begin{align}
a^\mathrm{short}_{ij}(r) = &a_{ij}(r) - {2\pi\over r^2}\int_0^r\mathscr{F}_{r'}^{-1}\left\{\bar{\phi}_{ij}(k)\exp(-k^2r_s^2)\right\}~\dd r' \nonumber \\
& + {2\pi\over r} \mathscr{F}_{r}^{-1}\left\{\bar{\phi}_{ij}(k)\exp(-k^2r_s^2)\right\}.
\label{eqn:short_accel}
\end{align}
To maintain requisite precision during the computation, the
integration in $r$-space is performed using a 4-point Newton-Cotes
formula, giving an error for the integration bounded by
\begin{align}
&< \left(\frac{r_s}{10N_\mathrm{TAB}\mathscr{O}}\right)^5 \partial_r^3 \mathscr{F}^{-1}_r\left\{\bar{\phi}_{ij}(k)\exp\left[-k^2r_s^2\right]\right\} \nonumber\\
&<\frac{r_s}{\left(10N_\mathrm{TAB}\mathscr{O}\right)^5} \label{eqn:accel_error}
\end{align}
where $\mathscr{O} \in \mathbb{N}$ is a user-specified parameter which
controls the resolution of the FFT.  The error bound in
Equation~\ref{eqn:accel_error} follows from application of standard
inequalities and our constraint of the $a_{ij}$.

\subsection{Limitations}
\label{sec:limitations}
Stock computes the non-periodic $k$-space Greens' function through
explicit FFT of the sampled $r$-space potential on a mesh of twice the
dimension desired for use in simulation.  Unfortunately,
generalization of this procedure from the specification of the
periodic $k$-space Greens' function as is done in the accurate
computation of the truncated short-range force is not practical due to
memory constraint.  Similarly, sampling the transformed radial
function enough to guarantee the requisite accuracy on all points of
the lattice through cubic interpolation encounters similar memory
constraint.  Since non-periodic computations may be performed to
unlimited dynamic range with the extended Tree algorithm, and since
large-scale investigations of modified force laws would naturally
proceed investigations on smaller scales, we do not believe this to be
a serious omission.

In \textsc{gadget}-2, collisional forces are only computed for one
type of ``gas'' particle.  As highlighted by
Marri \& White\cite{Marri21102003}, distinct gas species can allow the
effective capture of a broad range of physical phenomena involved in
galaxy formation and evolution.  While it is possible to alter the
gravitational force laws involving the gas, \emph{ngravs} does not
implement multiple gas species.  This choice was one of simplicity and
further extension to multiple gas species, i.e. following
Scannapieco~\emph{et. al.}\cite{Scannapieco21092006}, is
straightforward and could form the basis of future work.

At present, particle interactions under force laws with mass dependent
scale are only accurately computed for uniformly massed gravitational
species.  This applies to both active and passive mass dependence.
Without uniform masses, accurate computation is only possible only in
pure tree, non-periodic mode.  This is due to the Fourier
computation's use of a single momentum-space Greens' function for all
contributing densities, and to the necessary precomputation of force
and potential correction tables for direct infinite lattice
contributions.  Even in non-periodic pure tree mode, due to the
monopole averaging procedure, considerable force errors could result
given situations where a node contains comparable numbers of
same-species particles with different masses.  The efficient and
accurate computation of such interactions is an open question.

\section{Test problems}
\label{sec:testing}
Since \textsc{gadget}-2 has been well vetted over the past decade, to
verify the correctness of the \emph{ngravs} extension, it suffices to
demonstrate that \emph{ngravs} with Newtonian interactions and stock
agree in their common operational modes.  In addition to these
consistency checks, we also investigate the TreePM and Pure Tree
algorithms under more general force laws.

Profiling results presented in Table~\ref{tbl:runtimes} and
Figure~\ref{fig:Dvar_runtimes} were carried out to double precision on
a dedicated machine using 32 of 48 available cores for computation
with 94 gigabytes of RAM.  Periodic profiling and force accuracy runs
were performed at double precision with a 128 point mesh.  TreePM
transition studies were performed on a 256 point mesh.  Simulation
initial conditions are those included with stock, with particle type
reassigned as necessary to test \emph{ngravs}.  Our parameters can be
found in Appendix~\ref{sec:appendix}.

\subsection{Comparison with stock}
\begin{figure}[t]
\centering
\caption{\label{fig:ft_puretree} Fraction of forces computed during
  entire simulation with force error in excess of $\Delta f /f$, for
  Pure Tree mode and non-periodic boundary conditions, with $D=3$
  \emph{ngravs} (blue, dashed).  All interactions are Newtonian to
  permit direct comparison with stock (magenta, solid).  Residuals
  $|\Delta|$ from stock shown at top.  Note slightly improved force
  accuracy for \emph{ngravs} due to more detailed characterization of
  the mass distribution.}  
\includegraphics{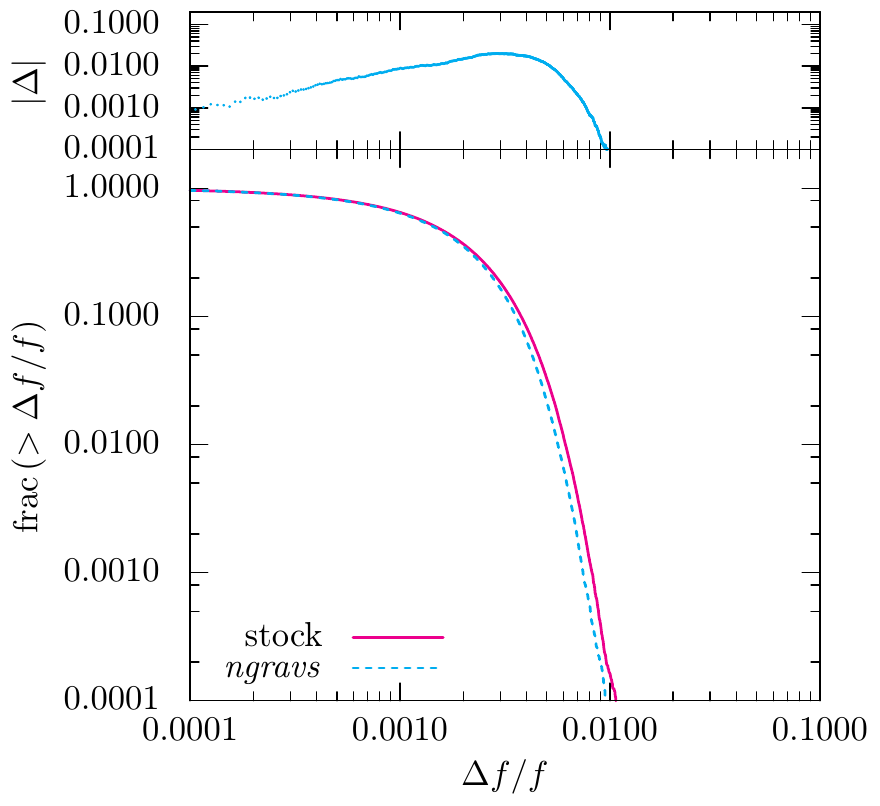}
\end{figure}%
\begin{figure}[t]
\centering
\caption{\label{fig:ft_treepm} Fraction of forces computed during
  entire simulation with force error in excess of $\Delta f /f$, for
  TreePM mode and periodic boundary conditions, with $D=3$
  \emph{ngravs} (blue, dashed).  All interactions are Newtonian to
  permit direct comparison with stock (magenta, solid).  Residuals
  $|\Delta|$ from stock shown at top.  Note that an erroneous
  softening scale of 600kpc in the default stock $\Lambda$CDM initial
  condition was reduced to 50kpc to keep the softening scale below the
  cutoff for use of the PM computation.}  
\includegraphics{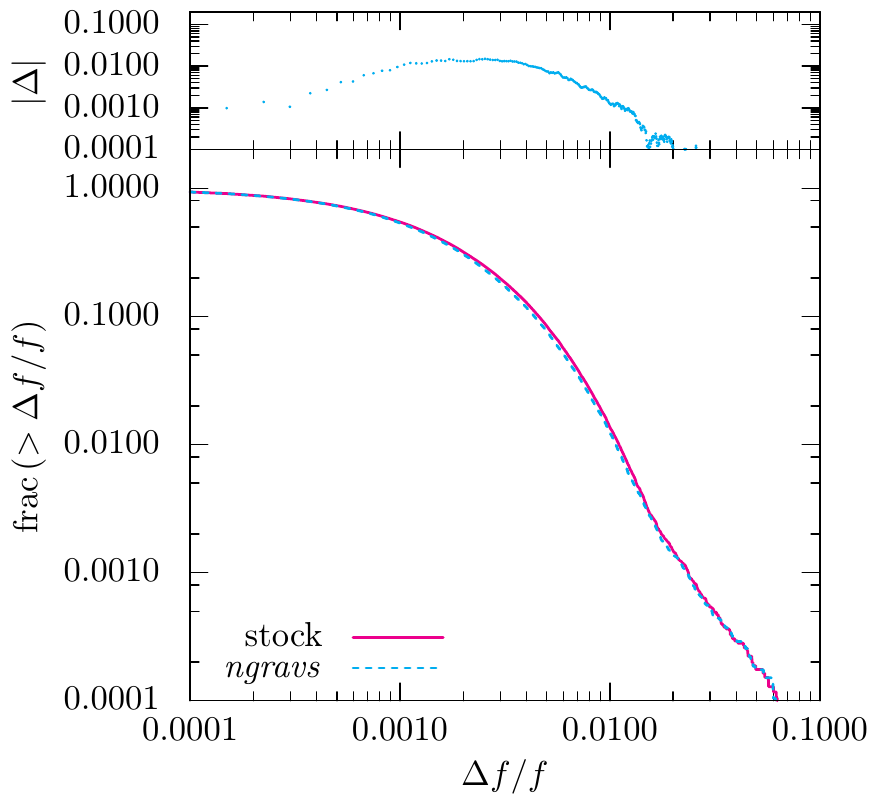}
\end{figure}%
We present force accuracy comparisons between stock and $D=3$
\emph{ngravs} for two of the four stock included test initial
conditions.  In Figure~\ref{fig:ft_puretree}, we demonstrate
consistency with stock for pure tree operation under non-periodic
boundary conditions.  In Figure~\ref{fig:ft_treepm}, we demonstrate
consistency with stock for TreePM operation under periodic boundary
conditions.  This test verifies the correctness of the lattice
summation, as the direct force computation requires the lattice
correction under periodic boundary conditions.  This periodic test
also includs a collisional ``gas'' species, which further verifies the
integrity of the unmodified collisional code.  All figures display the
fraction of force computations with force error
\begin{equation}
\Delta f / f \equiv \frac{\left|\vec{F}_\mathrm{alg} - \vec{F}_{N^2}\right|}{\left|\vec{F}_{N^2}\right|}
\end{equation}
in excess of that fraction.  Here, subscripts ``alg'' and $N^2$
represent computation by tree/mesh routines and direct Newtonian
summation, respectively.  The collapsing gas sphere was
excluded as the present implementation of \emph{ngravs} requires that
all collisional particles be of the same gravitational species.  Note
that the force accuracies are virtually identical; favorable residuals
indicate relatively minor improvements due to three multipole moments
per node.

\begin{figure}[t]
\centering
\caption{\label{fig:xition_coloumb} Force error as a function of
  separation for the Coloumb interaction with periodic boundaries.
  Stock is shown in red (grey), and \emph{ngravs} $D=1$ in blue
  (black).  Errors stacked from ten distinct simulations of a randomly
  placed massive source interacting with randomly placed test
  particles.  Vertical lines represent the mesh scale (dotted) and
  transition scale (dashed).  Binned rms shown at top, with $\alpha \equiv 0.005$
  the specified error tolerance for the tree algorithm.}
\includegraphics{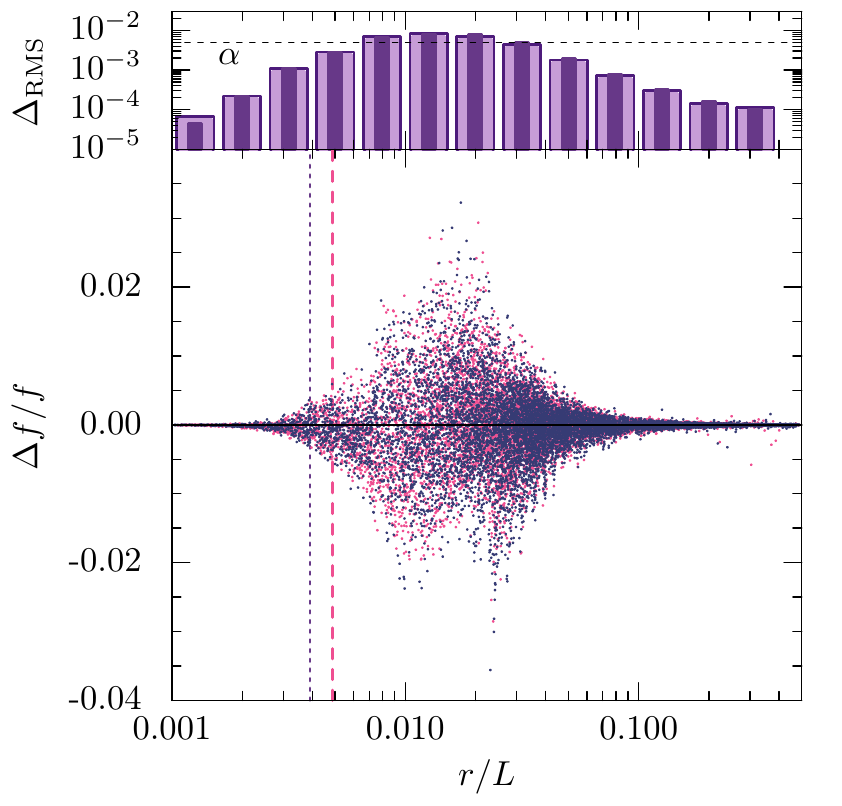}
\end{figure}%
In addition to characterizing errors across entire simulations, we
have also investigated the detailed error behavior of the TreePM
algorithm in the transition region.  We perform this test by creating
a sequence of initial conditions each with a randomly placed massive
source and shells of $\sim5000$ randomly placed test particles.  The
test particle density drops as $1/r^3$ so that the number of
interactions per shell is roughly constant.  All particles have zero
initial velocities and we consider only the first force computation,
recording force errors for all interactions.  The results of 10 of
these runs are then stacked.  Performance with the Newtonian
interaction can be seen in Figure~\ref{fig:xition_coloumb}, where the
force error behavior is indistinguishable from that of stock.

\subsection{The pure Yukawa interaction}
\begin{figure}[t]
\centering
\caption{\label{fig:xition_yukawa} Force error as a function of
  separation for the Yukawa interaction with periodic boundaries.
  Yukawa field masses $y_m = 10$ (blue, black) and $y_m = 50$ (pink,
  grey) shown.  Errors for each field mass stacked from ten distinct
  simulations between a randomly placed massive source, with randomly
  placed test particles.  Vertical lines represent the mesh scale
  (dotted) and transition scale (dashed).  Binned rms shown at top,
  with $\alpha\equiv 0.005$ the specified error tolerance for the tree
  algorithm.}  
\includegraphics{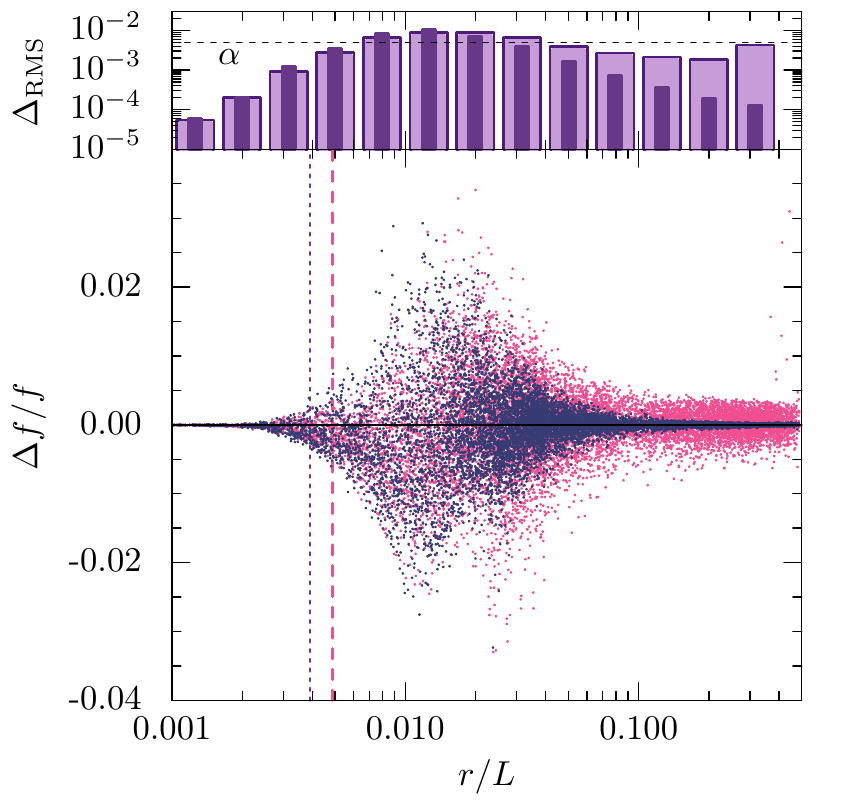}
\end{figure}%
To verify correct and accurate behavior of the TreePM transition
algorithm with more general forces, we have implemented the pure
Yukawa (Yukawa) interaction.  The Yukawa interaction represents a
pathological ``edge case'' for the $k$-space Gaussian low-pass filter
approach, because its $r$-space behavior is already exponentially
suppressed before filtering.  Conveniently, there exists significant
literature on lattice sums involving the Yukawa potential, and our
reference lattice implementation follows that of Salin and
Caillol\cite{Salin00}.  By comparison with a box filter, it was
determined that the tree contribution remained non-negligible relative
to the PM contribution under the Gaussian filter.  Fortunately, we
found that this pathology, specific to the Gaussian filter, can be
corrected by simply scaling the Yukawa $\vec{k}$-space Greens'
function by $\exp(-y_m^2r_s^2)$ where $y_m$ is the Yukawa field mass
and $r_s$ is the filter transition scale.

In Figure~\ref{fig:xition_yukawa} we demonstrate the error performance
for dimensionless $y_m\in\{10, 50\}$, with 50 sufficient to give
eventual $10\times$ suppression of the Coloumb potential over the
transition region.  Note that the rms error is essentially unchanged
throughout the transition region.  The rms error at large $r$
increases with increasing $y_m$ due to the presence of additive
exponential terms in the reference lattice sum and subsequent loss of
precision.  We note that this force test can be performed accurately
in \emph{ngravs} because the source and test masses can be assigned
distinct gravitational types.  Intra-type interactions can then be
turned off completely.  This is relevant for proper investigation of
the pure Yukawa interaction, as exponential suppression makes nearer
neighbors relevant, even with a very massive source.

In addition to pure Yukawa, we also have explored an evenly weighted
sum of Yukawa and Coloumb to verify that our correction factor to
Yukawa is robust.  We find that the error performance is essentially
unchanged from that of Coloumb, which verifies appropriate behavior
in the transition region of the summed force.

\subsection{The accumulator}
\begin{figure}[t]
\centering
\caption{\label{fig:ft_bam} 
Internal check of \emph{ngravs} $N_\perp$
  feature.  Force accuracy comparison in pure tree mode between two
  distinct gravitationally interacting species: one Newtonian, the
  other as described with $\beta \equiv 1.31\times10^{-6}$.  Note that
  force accuracy comparable to \textsc{gadget}-2 is maintained when
  the number of particles contributing to monopole approximations,
  $N_\perp$, is tracked (yellow, solid) and used to adjust the force.
  Without such tracking, the force errors increase by an order of
  magnitude (orange, dashed).}
\includegraphics{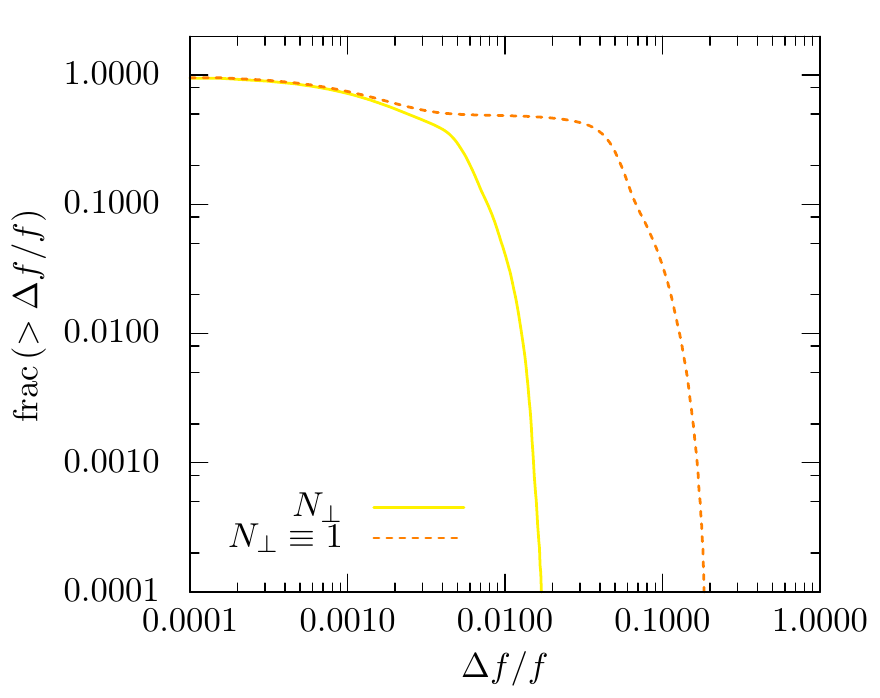}
\end{figure}

To verify the newly introduced optional tracking of contributing
particle counts $N_\perp$ in tree computations, we introduce a
gravitational species where $N_\perp$ can be used to give an exact
correction to the monopole approximation of the force laws.  This test
involves two species: species zero interacts via the usual Newtonian
interaction
\begin{align}
\Phi_{00}(r) &= -\frac{M_0 m_0}{r}
\end{align}
where we use uppercase to denote the active gravitational mass and
lowercase to denote the passive mass.  The second species, denoted
one, is characterized by a dimensionless scale $\beta$ and interacts
as
\begin{align}
\label{eqn:bambam}
\Phi_{11}(r) &= -\frac{2M_1 m_1}{\pi r}\tan^{-1}\left(\frac{4\pi\beta r}{M_1/N_\perp + m_1}\right).
\end{align}
Note that the interaction approaches Newton's at large $r$, but
softens to a constant as $r \to 0$.  This is the exact potential
between two hypothetical, spherically symmetric, cored densities of
the following form
\begin{align}
\rho(r, M) = \frac{M^2}{4\beta\pi^3}\frac{1}{\left(\left[M/2\beta\pi\right]^2 + r^2\right)^2}
\label{eqn:bam_density}
\end{align}
where $M$ is the total mass enclosed over all space by an object with
density given by Eqn.~\ref{eqn:bam_density}.  Interactions across
species are of the same functional form as Eqn.~\ref{eqn:bambam},
apart from the softening scale, which continues to be set by the cored
object
\begin{align}
\Phi_{10}(r) &= -\frac{2M_0 m_1}{\pi r}\tan^{-1}\left(\frac{4\pi\beta N_\perp r}{m_1}\right) \label{eqn:10} \\
\Phi_{01}(r) &= -\frac{2M_1 m_0}{\pi r}\tan^{-1}\left(\frac{4\pi\beta N_\perp r}{M_1}\right) \label{eqn:01}.
\end{align}
Note that Newton's third law is not violated, the distinct
Eqns.~\ref{eqn:10} and \ref{eqn:01} distinguish between the passive
and active mass for correct computation.  One may think of this
interaction as a preference for phenomenological cored
halos\cite{Walker09, de2009core} motivating a new class of
hypothetical object, the Massive Astronomical Halo (Mahalo).  The
results in Figure~\ref{fig:ft_bam} exhibit the force accuracy achieved
through this novel $N_\perp$ feature.

\section{Conclusion}
We have detailed the \emph{ngravs} extension to the massively parallel
hybrid Tree and mesh $N$-body code \textsc{gadget}-2, which now
permits consideration of $D^2$ gravitational interactions between $D$
particle species.  Periodic simulations can proceed in either Tree or
hybrid TreePM mode, while non-periodic simulations may be run in Tree
mode.  Our implementation vectorizes over the existing monopole
moments within the Barnes-Hut tree, and distinguishes between active
and passive gravitational mass during the mesh computations.  Memory
consumption remains favorable: particle data storage is unchanged from
\textsc{gadget}-2, tree storage is increased by $\sim 0.3(D-1)$, and
Fourier storage requirements are unchanged.  We subject the code to
numerous tests to gauge performance both in runtime and in force
accuracy.  We verify that runtime is dominated by tree performance and
scales as $\kappa D$ for $\kappa \in (1, 1.43)$ relative to that
  of \textsc{gadget}-2.  We find qualitatively identical,
  slightly improved, force accuracies compared to \textsc{gadget}-2:
  behavior expected from our particular implementation.  We also have
  introduced and verified a novel feature, which tracks the number of
  contributing particles of all species to any given monopole
  approximation, which can then be used to correct exotic force laws
  with dynamic softening lengths.  We believe that the \emph{ngravs}
  extension will facilitate investigation and constraint of exotic
  gravitational scenarios and have released \emph{ngravs}
  publicly\footnote{The website for downloading \emph{ngravs} is\hfill
    \\ \href{https://github.com/kcroker/Gadget-2.0.7-ngravs}{https://github.com/kcroker/Gadget-2.0.7-ngravs}}
  to the research community.

\section*{Acknowledgments}
The author would like to thank Naoki Yoshida and Junichi Yokoyama for
hospitality and encouragement, Volker Springel for warm and thorough
feedback, and Larry Glasser for the elegant integration procedure
leading to Eqn.~(\ref{eqn:bambam}).  The author additionally thanks Tom
Browder, Manuel Hohmann, and Brandon Wilson for useful discussions
during code review.  Significant portions of this work were performed
at the Kavli Institute for Physics and Mathematics of the Universe
(IPMU) and The University of Tokyo Research Center for the Early
Universe (RESCEU) under joint support from National Science Foundation (NSF)
Grant 1415111 and the Japan Society for the Promotion of Science (JSPS).
Additional work was performed at the University of Tartu Institute of
Physics, supported by the US Department of State under a Fulbright
Student Award.

\appendix
\section{Initial conditions and simulation parameters}
\label{sec:appendix}
For completeness, we characterize the initial conditions and paired
simulation parameters packaged with \textsc{gadget}-2, which we have
used to test and benchmark \emph{ngravs}.  These initial conditions
were chosen for convenience and consistency, and they exercise the
full range of modified functionality within \emph{ngravs}.  For
simulations involving $D>1$, collisionless particles were distributed
evenly into the remaining types and assigned identical parameters.
\begin{table}
\caption{\label{tbl:simulation_params}Initial conditions and
  parameters for the Pure Tree test case. $\epsilon$ is the softening
  length.}
\begin{center}
{\renewcommand{\arraystretch}{1.1}
Collision of 2 spiral galaxies \\
\begin{tabular}{r|l}
\hline
$|x_i|$ & $< 200$~kpc \\
$|v_i|$ & $<350$~km/s \\\hline
Type \#1 & Collisionless \\
& $N=4\times 10^4$ \\ 
& $m=1.05\times 10^{-3}~M_\mathrm{\astrosun}$ \\ 
& $\epsilon=1$ \\\hline
Type \#2 & Collisionless \\
& $N=2\times 10^4$ \\ 
& $m=2.33\times 10^{-4}~M_\mathrm{\astrosun}$ \\ 
& $\epsilon=0.4$ \\
\hline
\end{tabular}}
\end{center}
\end{table}%
\begin{table}
\caption{\label{tbl:simulation_params}Initial conditions and
  parameters for the TreePM test case.  Cosmological fractions and
  Hubble parameter are the standard values.  $\epsilon$ is both the
  comoving and physical softening length, which was changed from an
  erroneous default value that caused the softening scale to
  overlap the PM transition region.}
\begin{center}
{\renewcommand{\arraystretch}{1.1}
$\Lambda$CDM Universe from $z=10$ \\
\begin{tabular}{r|l}
\hline
$|x_i|$ & $< 5\times 10^4$~kpc \\
$|v_i|$ & $<10^3$~km/s \\\hline
Type \#1 & Collisional \\
& $N=2^{15}$ \\ 
& $m=4.24~M_\mathrm{\astrosun}$ \\ 
& $\epsilon=50$ \\\hline
Type \#2 & Collisionless \\
& $N=2^{15}$ \\ 
& $m=27.5~M_\mathrm{\astrosun}$ \\ 
& $\epsilon=50$ \\
\hline
\end{tabular}}
\end{center}
\end{table}%
\begin{table}
\caption{\label{tbl:simulation_params}Initial conditions and
  parameters for the gas collapse profiling run.  Note that masses are
  specified per particle in this initial condition.}
\begin{center}
{\renewcommand{\arraystretch}{1.1}
Collapsing gas sphere \\
\begin{tabular}{r|l}
\hline
$|x_i|$ & 0.1~cm \\
$|v_i|$ & $=0$ \\\hline
Type \#1 & Collisional \\
& $N=1472$ \\ 
& $m_i=6.79\times 10^{-4}$~g \\ 
& $\epsilon=0.004$ \\
\hline
\end{tabular}}
\end{center}
\end{table}%

\bibliography{ngravs2}

\end{document}